\documentclass[pre,aps,preprint,showpacs]{revtex4}
\usepackage{revsymb}
\usepackage{graphicx,psfrag}
\usepackage{amsmath}
\begin{document}
\author{Steffen Trimper, Knud Zabrocki}
\affiliation{Fachbereich Physik,
Martin-Luther-Universit\"at,D-06099 Halle Germany}
\email{trimper@physik.uni-halle.de}
\title{Delay-Controlled Reactions}
\date{\today }

\begin{abstract}
When the entities undergoing a chemical reaction are not available 
simultaneously, the classical rate equation of a reaction or, 
alternatively for the evolution of a population, should be extended by 
including non-Markovian memory effects. We consider the two cases of an 
external feedback, realized by fixed functions and an internal feedback 
originated in a self-organized manner by the relevant concentration itself. 
Whereas in the first case the fixed points are not changed, although the 
dynamical process is altered, the second case offers a complete new behaviour,  
characterized by the existence of a time persistent solution. 
Due to the feedback the reaction may lead to a finite concentration in 
the stationary limit even in case of a single-species pair annihilation 
$A + A \to 0$ process. We argue that the different cases are similar to a coupling of 
additive or multiplicative noises in stochastic processes.

\end{abstract}
\maketitle

\section{Introduction}

Chemical reactions involving different species are described by 
classical rate equations in which the time evolution for the forward and reverse 
reactions are balanced by the product of the concentration of the reacting entities, 
see for a recent review \cite{0}. The approach is based on the assumption that 
the reactants are available simultaneously. In the one species-annihilation process 
a particle $A$ is annihilated upon encounter according to the reaction scheme 
$A + A \to 0$. The reaction is immediate and is realized with a certain rate $\nu$. 
During the related one-species coalescence process $A + A \to A$ the reactants fuse 
together realized with another rate $\mu$. The time evolution of the global 
concentration $c(t)$ obeys the mean field equation 
\begin{equation}
\frac{dc(t)}{dt} = -(2 \nu - \mu) c^2
\label{ev}
\end{equation}
whereas $\nu$ and $\mu$ are the coalescence and annihilation rates. In case 
the reaction takes place only after a sufficient accumulation of the reactants, 
the reaction process is changed crucially. We demonstrate that the 
long-time behaviour is typically dominated by such delay effects. The system is able to reach a 
stationary state instead of exhibiting an algebraic decay in time as it follows 
using Eq.~(\ref{ev}). Let us illustrate the situation in mind in more detail. 
Generally the time evolution of the concentration $c(t)$ is characterized by 
gain and loss terms. Moreover, the concentration could also depend on the history 
of the sample to which it belongs. In the same sense the changing rate of the 
concentration should be influenced by the changing rate in the past. Thus the 
evolution for $c(t)$ has to be supplemented by a memory term indicating the 
non-Markovian behaviour. Such a term models, for instance the way on which a seed 
concentration had been accumulated by a delayed transport mechanism originated 
by the environment of the reactants. With other words, the changing rate of a certain 
quantity at time $t$ is also determined by the accumulation rate at a former time 
$t^{\prime} < t$. In between, i.e. within the interval $\tau = t - t^{\prime}$, the 
reactants are enriched while changing the concentration at $t^{\prime}$. Regardless 
that process and further fluctuations the available amount of concentration 
at time $t$ is governed by instantaneous gain and loss of concentration as well as on 
the changing rate at former times $t^{\prime}$. Consequently the evolution  
Eq.~(\ref{ev}) should be modified according to 
\begin{equation}
\partial_t c(t) = r c(t) - u c^2(t) - \int_0^t K(t,\, t^{\prime};c)
\partial_{t^{\prime}}c(t^{\prime}) dt^{\prime}\quad .
\label{ev1}
\end{equation}
The last term represents a permanent re-evaluation of the reactions. Hence the aim of the present 
paper is to study models which follow an evolution equation as given by Eq.~(\ref{ev1}) where 
the memory kernel $K$ will be specified below. As a new ingredient we assume 
that the memory term may depend on the concentration $c(t)$ and its derivative as 
already indicated in Eq.~(\ref{ev1}). Whereas the first realization consists of a kernel, 
given by a deterministic function, the second case is characterized by a kernel, determined by 
the concentration itself. The first case is denoted as an external feedback with a distributed 
time-delay, the second one is signified as an internal delay, since the time scale of the kernel is 
defined in a self-organized manner by the time scale of the concentration itself.\\ 
Our model can be grouped into the increasing interest of incorporating delay and feedback 
mechanism in a large variety of systems. Obviously the crucial factor governing the 
dynamics of systems comprising many "units" consists of interaction and competition. 
Such features are believed to underlie the complex dynamics observed in disciplines as 
diverse as economics \cite{2a}, biology and weather \cite{3}, politics \cite{4}, 
medical care \cite{5}, and ecology \cite{6}. As discussed recently and also stressed by 
the present paper another characteristic trait of physical \cite{6a,6b} as well as biological 
systems \cite{6c} is played by the mentioned time-delayed couplings. Such memory effects, 
considered in a large variety of systems, could be a further unifying feature of complex 
systems. It is well-known that evolution equations with memory kernels as it is presented 
by Eq.~(\ref{ev1}), can be derived following the well established projector formalism 
due to \cite{mori}, see also \cite{naka}. That approach had been applied successfully for 
the density-density correlation function in studying the processes in undercooled 
liquids \cite{l,goe}. Recently a Fokker-Planck equation with a non-linear memory term 
was used to discuss anomalous diffusion in disordered systems \cite{ss}. The results could 
be confirmed by numerical simulations including diffusion on fractals \cite{bst}, 
see also \cite{t,mo}. Moreover, it had been demonstrated \cite{loc} that mobile particles 
remain localized \cite{loc} due to the feedback-coupling. Notice that a formal solution of 
whole class of non-Markovian Fokker-Planck equations can be expressed through the solution 
of the Markovian equation with the same Fokker-Planck operator \cite{s}. The non-Gaussian 
fluctuations of the asset price can be also traced back to memory effects \cite{st}. 
An additional cumulative feedback coupling within the Lotka-Volterra model, which may stem 
from mutations of the species or a climate change, leads to a complete different 
behaviour \cite{zab} compared to the conventional model. If the Ginzburg-Landau model 
for the time evolution of an order parameter is supplemented by a competing memory term,  
the asymptotic behaviour and the phase diagram is completely dominated by such a term 
\cite{zab1}. Different to our self organized approach with $K(t,t^{\prime}; c)$ there 
is a broad class of models with delay integrals without a dependence on the variable $c(t)$ 
\cite{oy,ge,fe}, for a survey and applications in biology see \cite{mur}. 
The spreading of an agent in a medium with long-time memory, which can model epidemics, is 
studied in \cite{gr}. Time-delayed feedback control is an efficient method for stabilizing 
unstable periodic orbits of chaotic systems \cite{p}. It may also induce various 
patterns including travelling rolls, spirals and other patterns \cite{j}. A further 
approach concerns the study of bistable time-delayed feedback systems driven by noise 
\cite{m}. Based on a former paper \cite{tp} the distribution function of the 
first-passage-time is analysed where both, the delay and the noise term offers a 
significant influence on the behaviour. A global feedback is studied recently also in 
a bistable system \cite{sa}. The purpose of that paper is a discussion of the 
domain-size control by a feedback. Even in an 
open quantum system non-Markovian dynamics is characterized by a time-non-locality in 
the equation of motion for the reduced density operator \cite{mm}.\\ 
In view of the large variety of systems with feedback couplings it seems to be worth to study 
simple models, which still conserve the crucial dynamical features of evolution models as 
non-linearities and moreover as a new ingredient, delayed feedback-coupling. In the 
present paper we demonstrate the crucial influence of the non-Markovian memory term on 
chemical reactions, where the retardation effects are comprised into a memory kernel $K(t)$. 
That memory kernel yields an additional competitive term to the instantaneous non-linear terms.

\section{External Feedback}
\subsection{Discrete time-delay}

In this section we specify the model, Eq.~(\ref{ev1}), by fixing 
the memory kernel $K$ in terms of deterministic functions. This procedure is well-known in the 
framework of population dynamics (an introduction is given in \cite{banks}). 
Different to these approach our rate equation (\ref{ev1}) offers a feedback-coupling 
to the decay rate of the population or the density of a chemical species $\partial_t c(t)$, 
respectively. By choosing the kernel $K(t,t^\prime ;\tau )=\mu \, \delta (t- t^\prime -\tau)$, 
and substituting $K$ in the dimensionless form of Eq. (\ref{ev1}), one gets a discrete 
time-delayed differential equation
\begin{equation}\label{disdel}
\frac{dc(t)}{dt} = c(t)- c^2(t) -\mu \, \frac{dc(t-\tau )}{dt} \quad .
\end{equation}
Here the quantity $\tau $ is the delay time which is assumed to be short ranged $\tau \ll t$. 
If $\tau =0$, it results in a modification of  the evolution equation for the logistic 
growth, which is solvable (see below). The limiting case $\tau \rightarrow 0$ can be obtained 
by expanding the delayed term of Eq. (\ref{disdel}). This leads to an equation  
\begin{equation}
\frac{d^2 c}{dt^2}-\frac{1+\mu}{\mu\, \tau} \,\frac{dc}{dt} +\frac{1}{\mu \, \tau} c\, [1- c]=0 
\qquad \mu \neq 0,
\end{equation}
which  cannot be solved in closed form due to the non-linearity. One gets an analytical 
solution, if one neglects the non-linearity, which corresponds to the exponential 
growth of the particle density in addition of a feedback-coupling. Using standard methods 
the solution $c (t)$ is given by
\begin{equation}\label{disdel2}
c(t) = C_1 \, \exp (\kappa_{+}\, t )+ C_2 \, \exp (\kappa_{-} \,t )\quad \text{with}\quad \kappa_{\pm}=
\frac{1}{2\,\mu\, \tau}\left( 1 + \mu  \pm \sqrt{(1+\mu )^2 -4\,\mu\,\tau } \right)\quad .
\end{equation} 
The integration constants can be calculated to 
$C_1 =c_0 \, \frac{\kappa_{-}-1}{\kappa_{-}-\kappa_{+}}$ and 
$C_2 =c_0 \, \frac{1-\kappa_{+}}{\kappa_{-}-\kappa_{+}}$, if one suggests $c(0)=c_0$ and $\frac{dc}{dt}=c_0$. 
Discussing the stability of Eq. (\ref{disdel2}) one finds out, that for $\mu >0$ the both exponents 
$\kappa_{\pm}$ are positive. In case of $\mu <0$ one of them is positive, i. e. there are no bounded 
solution in the limit $t\rightarrow \infty$. Alternatively, one can treat the differential equation with delay 
by using the Laplace transformation. Denoting the transformed function by 
$c(z) \equiv \mathcal{L}\{c (t)\} = \int_0^{\infty} c(t) \exp(-zt)dt$ the transformed function of the linear 
equivalent of Eq.~(\ref{disdel}) obeys the relation
\begin{equation}\label{disdel3}
c(z)=\frac{c_0\, [1+(1+\mu)\,\exp(-\tau\, z)]-c(-\tau)+z\,\exp(-\tau\, z)\,
\int\limits_{-\tau}^{0} \exp(-\tau\, \xi)\, c(\xi )\, d\xi}{z\, [1+\mu \,\exp(-\tau\, z)]-1}\quad .
\end{equation} 
As displayed by the last relation, differential equations with delayed terms are distinguished from 
conventional equations by one important feature, that the function $c(t)$ has to be given within 
the whole interval $-\tau < t < 0$ and not only at one special point $c (t=0) =c_0$. Fixing $c (t) =c_0$ in that 
interval $-\tau < t <0$, then the Eq.~(\ref{disdel3}) becomes simpler
\begin{equation}\label{disdel4}
c (z)=\frac{c_0}{z-\frac{1}{1+\mu\,\exp(-\tau\, z)}}\quad .
\end{equation}
This equation has to be transformed back. Since the zeros of the denominator of Eq.~(\ref{disdel4}) are not 
available analytically, the function $c(t)$ could not be not calculated exactly. However, one can convince that 
in the long time limit, $z \, \tau \ll 1$, the concentration $c(t)$ offers an exponential form, already obtained in 
Eq.~(\ref{disdel2}). 

\subsection{Distributed time-delay - Exponential kernel}

\begin{figure} [!ht]
\centering
\psfrag{la}{$\lambda$}
\psfrag{p0}{$c_0$}
\psfrag{mu}{$\mu$}
\includegraphics{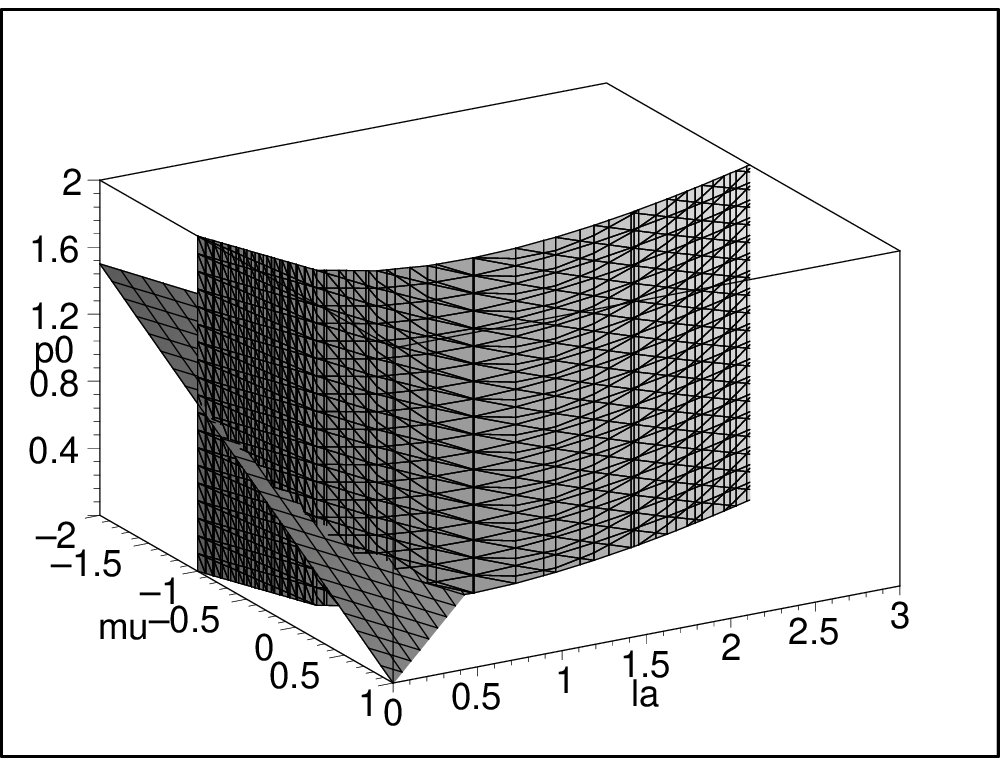}
\caption{Stability domain for an exponential kernel in the $(\mu ,\lambda , c_0 )$ plane.} 
\label{Fig.1}
\end{figure}

An usual choice of a kernel in problems dealing with continuously distributed time-delay is the exponential
kernel $K (t,t^\prime )=\mu\, \exp \left[-\lambda\, (t-t^\prime )\right]$ ($\lambda >0$) , 
which satisfies some properties like the boundedness and positivity of the kernel. The parameter 
$\lambda $ determines the time scale of the memory. Substituting this kernel in 
dimensionless form of Eq. ~(\ref{ev1}) it leads to the following equation
\begin{equation}\label{disdel5}
\partial_t c(t) = c(t) - c^2 (t) -\mu\, \int\limits_{0}^{t} 
\exp \left[-\lambda\, (t-t^\prime )\right]\, \partial_{t^\prime} c(t^\prime)\, dt^\prime\qquad 
c(0) = c_0\quad.
\end{equation}
Fixed points of this equations are $c_s =0$ and $c_s =1$. They are independent of the memory parameter 
$\mu $ which is a general result, compare the discussion at the end of this section. To gain information  
about their stability, we investigate the linear equation in analogy to Eq. ~(\ref{disdel5}) firstly.
Using Laplace transformation results in
\begin{equation}\label{disdel6}
c (z) = c_0 \, \frac{z+\lambda +\mu}{(z-1)\, (z+\lambda ) +\mu\, z}\quad .
\end{equation}
The zeros of the denominator of the latter formula are given by 
$z_{\pm}= 1/2 \, [-(\lambda +\mu -1 ) \pm \sqrt{D}]$
with $D = (\lambda +\mu -1)^2 + 4\, \lambda$. It is obvious that $D>0$, because the 
assumption $\lambda >0$. Further one observes that $\sqrt{D}>|\lambda +\mu -1 |$ and so 
$z_+$ is positive and $z_{-}$ is negative. Because the structure of the 
solution is $c (t)= A\,\exp (z_+ \, t) + B\,\exp (z_- \, t)$ , we found only unstable 
or unbounded solutions (for $t\rightarrow \infty$), respectively.
That means that the exponential growth, which is observed in case of $\mu =0$, could 
not be restricted through the additional feedback term. Alternatively, Eq.~(\ref{disdel5}) can be 
analysed by a repeated differentiating with respect to $t$ and a renewed use of the first order equation 
to eliminate the integral. The procedure results in the second order equation
\begin{equation}\label{disexp}
\frac{d^2\, c}{dt^2} + [2\, c  +\lambda +\mu -1]\, \frac{dc}{dt} +\lambda\, c \, [c-1 ]=0 
\quad \text{with} \quad   c(0) = c_0 \quad \text{and} \quad \frac{dc(0)}{dt}= c_0\, [1- c_0 ]
\end{equation}
that linear equivalent exhibits the same behaviour found above. The non-linear 
Eq.~(\ref{disexp}) gives rise to a more complex stability behaviour.
To get insight into the stability behaviour of both fixed
points we take the ansatz $c (t) = c_s +\varphi (t)$. Substituting this ansatz into 
Eq.~(\ref{disdel5}) and omitting all terms of order $\varphi^2$ one gets the equations 
for $\varphi (t)$
\begin{equation}
\dot{\varphi} (t)=\pm\, \varphi (t) -\mu\, \int\limits_{0}^{t} 
\exp \left[-\lambda\, (t-t^\prime )\right]\, \dot{\varphi} (t^\prime )\, dt^\prime \quad ,
\end{equation} 
where the positive sign corresponds to $c_s = 0$ and the negative one to $c_s = 1$. The 
equation for the positive sign is already discussed above, and it follows that the trivial solution 
$c_s =0$ is an unstable fixed point. To gain the stability behaviour of the second 
fixed point we follow the same steps as above. The Laplace transform of  
$\varphi (t)$ is denoted as $\varphi (z) =\mathcal{L} \{\varphi (t)\} (z) = 
\int_0^{\infty} \varphi(t) \exp(-zt)dt $. We get
\begin{equation}
\varphi (z) = \varphi_0 \, \frac{z+\lambda +\mu}{(z+1)\, (z+\lambda ) +\mu\, z}\quad.
\end{equation}
In this case the zeros of the denominator are given by 
$1/2\, [-(1+\lambda +\mu )\pm \sqrt{D}]$, where
the discriminate is $D = (1+\lambda +\mu)^2 -4\, \lambda$. The solution of 
$\varphi (t)$ (for $D\not= 0$) can be written as   
$\varphi (t) = \varphi_0 \, \left[A\, \exp (z_+ \, t)+B\, \exp(z_- \, t)\right] $ 
with the coefficients
\begin{equation}
A=\frac{-1+\lambda +\mu +\sqrt{D}}{2\,\sqrt{D}}\quad \text{and}\quad 
B=\frac{1-(\lambda +\mu )+\sqrt{D}}{2\,\sqrt{D}}\quad.
\end{equation}  
The last formula implies that for $D\ge 0$ the solution is real and hence physically relevant.  
The limiting case $D=0$ delivers critical values either for $\mu$ or for $\lambda$. 
Choosing $\mu =\mu (\lambda )$ the critical values for $\mu$ are 
$\mu_{\pm}^c (\lambda )=-(\lambda +1 )\pm 2\, \sqrt{\lambda}$. If $\mu$ is situated 
in the intervals $\mu >\mu_{+}^c$ or $\mu < \mu_{-}^c$, $D>0$ is fulfilled and the 
resulting solution is real. The solution for $\varphi$ is only stable, if 
$\mu > -(1+\lambda )$. All together the parameter $\mu$ has to fulfil 
$\mu > \mu_{+}^c$ to get stable, physically relevant solutions. In the limiting case, 
$\mu =\mu_{+}^c$, one specifies
\begin{equation}
\varphi (t) =\varphi_0 \,\exp \left(-\sqrt{\lambda}\, t\right)\,\left[1+ (\sqrt{\lambda} - 1)\, 
t\right]\quad ,
\end{equation}
which is a bounded solution for $t\rightarrow \infty$. In case of 
$\lambda <1$, $\varphi (t)$ changes its sign at $t^\ast =(1-\sqrt{\lambda})^{-1}$.
Together we can determine the stability domain for the fixed point $c_s =1$ to the 
area in the $(\lambda,\, \mu)$-plane, where $\mu \ge \mu_{+}^c (\lambda )$ is fulfilled. 
Eq.~(\ref{disexp}) is an evolution equation, which is well-known as equation for a damped 
oscillator. Interpreting the equation as an equation of motion for a particle in a 
potential $U(c)= -c^2 /2 + c^3/3$, then the factor in front of the first derivative 
has the meaning of the damping constant. Due to the non-linearity of the underlying 
evolution equation the damping parameter $\gamma (c) = 2\, c (t) + \lambda +\mu -1$ is 
driven by the time dependent concentration. If one regognizes that only a positive 
damping parameter $\gamma (c) \geq 0$ is reasonable, it is obvious that the stability 
criteria is additionally depended on the initial value $c_0$. That is indeed a feature of systems 
including feedback or memory effects. With the following constraints  
\begin{eqnarray}
\mu + \lambda +1 -2\,\sqrt{\lambda} &\ge 0&\nonumber\\
2\, c_0 +\mu +\lambda -1 &\ge& 0
\end{eqnarray}
one gets the stability domain depicted in Fig.\ref{Fig.1}, where the $(\mu ,\lambda , c_0 )$- plane is 
displayed. The area, limited by the curved area and the plane area at the forefront, is the region 
where the solution is supposed to be stable.

\subsection{Distributed time-delay - a periodic kernel}

\begin{figure} [!ht]
\centering
\includegraphics{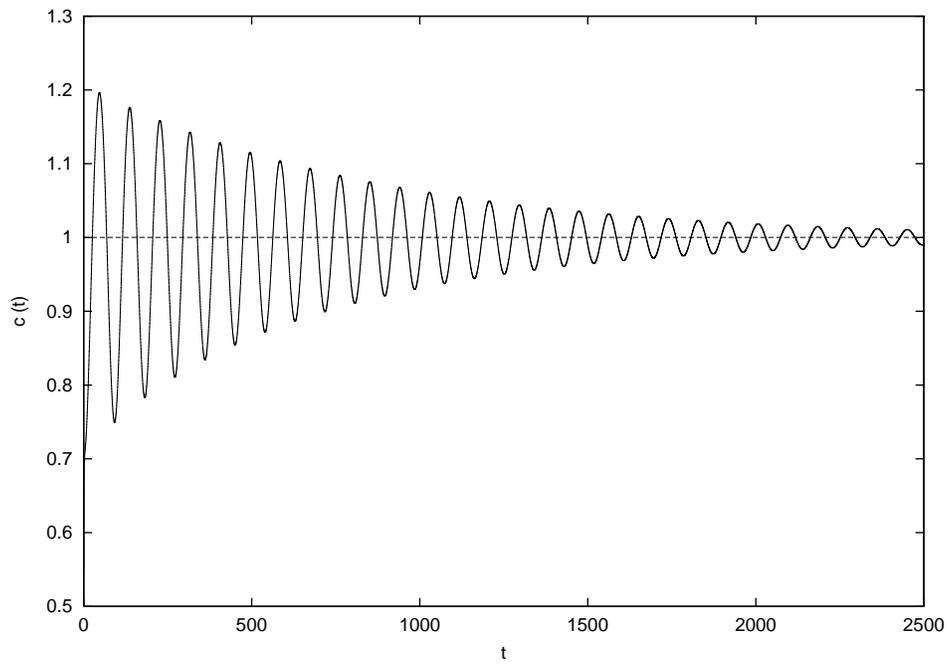}
\caption{Time evolution of $c(t)$ with a periodic kernel and $\lambda = 0.1, \mu = 1$ and $c_0 = 0.5$.}
\label{Fig.2}
\end{figure}

Since the dynamical behaviour is affected by the sign of the memory parameter $\mu $, we consider now 
a periodic kernel such as $K (t,\, t^\prime )=\mu\, \cos \left[\lambda\, (t-t^\prime )\right]$, which
results in
\begin{equation}\label{periodic1}
\partial_t c(t)= c(t)-c^2 (t) -\mu\, \int\limits_{0}^{t} \cos \left[\lambda\, (t-t^\prime )\right]\, 
\partial_{t^{\prime}} c(t^\prime )\, dt^\prime \quad .
\end{equation} 
Again the two stationary values are determined to $c_s =0$ and $c_s =1$. The 
Laplace transform  of the corresponding linear evolution equation is calculated to
\begin{equation}\label{periodic2}
c (z)= c_0 \, \frac{z^2 +\lambda^2 +\mu\, z}{(z-1)\, (z^2 +\lambda^2) +\mu\, z^2}\quad.
\end{equation} 
The denominator is a cubic polynom and the zeros could be calculated. At least one of the three solutions 
has a positive real part in case of $\lambda \not= 0$ and so there are only
unbounded solutions for the linear equivalent of Eq.~(\ref{periodic1}) for long times. Like in  the case
of the exponential kernel Eq.~(\ref{periodic1}) can be written as a conventional differential equation which is  
in the present case third order
\begin{equation}\label{periodic3}
\frac{d^3 c}{dt^3}+\left[2\, c+\mu -1\right]\, \frac{d^2 c}{dt^2}+\left[2\, 
\frac{dc}{dt}+\lambda^2  \right]\, 
\frac{dc}{dt}+\lambda^2 \, c\, (c-1)=0
\end{equation}
with the initial value $c(0)=c_0$, the initial slope $\frac{dc(0)}{dt}= c_0\, (1-c_0 )$ and the 
initial curvature $\frac{d^2 c(0)}{dt^2} = c_0 \, (1-c_0 )\, (1-\mu -2\, c_0 )$. The invariance of the kernel 
against the inversion of $\lambda \leftrightarrow -\lambda$ is maintained in Eq.~(\ref{periodic3}), too.
The third order differential equation is again equivalent to a system of three first order equations. The 
linear form leads to the same results as discussed after Eq.~(\ref{periodic2}). 
In Fig.\ref{Fig.2} we show a typical example for the time evolution of the concentration. Obviously, an increasing 
value of the parameter $\lambda $ leads to an overdamped behaviour.\\ 
The special case of a constant memory kernel $K (t,\, t^\prime ) = \mu$ can be treated exactly. 
Setting $\lambda =0$ in Eq.~(\ref{periodic1}) it results the following equation
\begin{equation}\label{periodic4}
\frac{dc}{dt}=[1-\mu ]\,  c(t) - c^2 (t) +\mu \, c_0 \quad .
\end{equation}
The decay of the concentration decay depends directly on the initial value $c_0$ making the memory 
apparently. Further the growth rate is modified by replacing $1 \to 1-\mu $. The general solution for 
arbitrary $\mu$ is deducible with the result 
\begin{align}\label{periodic6}
c (t) &=  \frac{1-\mu}{2} + \frac{\hat{D}}{2}\, \tanh\left[ \frac{1}{2}\, \hat{D} \, t  + \hat{c} \right] \nonumber\\
\mbox{with}\quad \hat{D} &=  \sqrt{(1-\mu )^2 + 4\, \mu \, c_0}\quad \mbox{and}\quad  
\hat{c} = \frac{1}{2} \ln \left[\frac{\hat{D} + 2c_0 + \mu -1}{\hat{D} - 2c_0 - \mu + 1}\right].    
\end{align}
The last solution is only reasonable within the interval $0< c_0 <1$. The special case 
$\mu =1$ yields
\begin{equation}\label{periodic5}
c (t)= \sqrt{c_0} \, \tanh \left[\sqrt{c_0}\, t +\frac{1}{2}\, \ln \left(\frac{1 + \sqrt{c_0}}{1 - \sqrt{c_0}}\right)\right]\quad ,
\end{equation}
The characteristic width of the curve $\tau _0 = 2 \hat{D}^{-1}$ is determined by the memory parameter $\mu$ and 
the initial value $c_0$. For $t \to \infty$ the solution, given by Eqs.~(\ref{periodic6}, \ref{periodic5}), 
lead to fixed points 
\begin{equation*}
c_s (\mu ) = \frac{1}{2}\,[1 - \mu \pm \hat{D}]
\end{equation*}
different from $c_s = 0$ or $c_s = 1$, respectively. The negative branch is omitted, because $c_s$ has to be positive.  
Contrary to the previous cases the 
stationary solution depends on the memory parameter $\mu $. Concerning the non-trivial case, where $\lambda \neq 0$, 
the instability of $c_s = 0$ is maintained for all values of $\mu $, whereas $c_s = 1$ is stable in the half-plane 
$\mu > 0  $. In this area one gets a solution starting with an exponential decay (growth) which is followed by a 
damped oscillational motion toward the stationary solution $c_s = 1$, compare Fig.\ref{Fig.2}.\\

The analysis can be generalized to an arbitrary kernel $K(t)$. Considering the 
Laplace transformed of Eq.~(\ref{ev1}), one concludes straightforwardly, that 
the fixed points remain unchanged, whenever the transformed memory kernel satisfies the relation 
\begin{equation}
\lim_{z \to 0} z\, K(z) = 0\quad .
\label{lt}
\end{equation}
For a constant kernel that equation is not fulfilled and one finds the modified fixed point given above. 
Thus, an external non-trivial feedback-kernel behaves  
like an additive noise in stochastic processes. In the next section we demonstrate how that picture is 
altered by assuming an internal feedback. In that case the kernel is determined by concentration itself. 
As consequence the last relation is likewise violated and the stationary points are directly controlled 
by the memory. Such a situation is the feature of stochastic processes whenever a multiplicative noise is 
included. 

\section{Internal Feedback}

In this section let us generalize the previous approach by including self-organized 
delay effects where, as already discussed in the introduction and in accordance with the conclusion of 
the last section, the kernel depends on the concentration $c(t)$ itself. That means, the time scale of 
$K(t)$ in Eq.~(\ref{ev1}) is determined by the time scale of $c(t)$. Following this line 
that the delay effects are dominated by the concentration $c(t)$ we conclude    
\begin{equation}
K(t,\,t^{\prime}; c) \equiv K\left(c(t - t^{\prime})\right)\quad.
\label{ev2}
\end{equation}
The kernel is defined by the concentration in the intermediate interval $t - t^{\prime}$ 
which is coupled to the changing rate at the previous time $t'$ as indicated in Eq.~(\ref{ev1}). 
The memory kernel characterizes the way on which a seed concentration had been accumulated. Within the 
time interval $\tau = t - t^{\prime}$, the concentration is further enriched, or with other words, 
at time $t^{\prime}$, only an incomplete reaction is realized. During the time interval 
$t - t^{\prime}$ the residual particles are moved to the reaction zone. 
As the simplest realization of Eq.~(\ref{ev2}) we chose a linear dependence on the concentration. 
In that case the memory term is a competitive one to the conventional quadratic term.    
Summarizing all contributions we will analyse the evolution equation
\begin{equation}
\partial_t c(t) = r c(t) - u c^2(t) - 
\kappa  \int_0^t c(t - t^{\prime})~\partial_{t^{\prime}} c(t^{\prime}) dt^{\prime}\quad.
\label{ev4}
\end{equation}
The first term characterizes the spontaneous creation of particles (with rate $r>0$) in 
according to the reaction $0 \to A$, whereas the second one describes a two particle 
reaction discussed in the introduction, compare Eq.~(\ref{ev}). 
Both terms leads to a stable stationary solution when they are balanced, 
manifested by $r > 0,~u > 0$. For generality one could also discuss the case that the 
parameter $r$ and $u$ can vary and change their signs. Here, we restrict us to positive quantities 
$r$ and $u$. The last term models the feedback with the coupling parameter 
$\kappa $ which mimics the influence of the environment. The following results are strongly 
influenced by the sign of the memory strength $\kappa$. In case of $\kappa < 0$ and 
$u > 0$  both non-linear terms are competitive ones. Let us remark that the memory kernel 
gives rise to a coupling of the time scales. In the vicinity of the upper limit of the integral 
$t^{\prime} \simeq t$ the memory term reads $c(0)\partial_t c(t)$, i.e. a momentary change at the 
observation time $t$ is coupled to the value at the initial time $t = 0$. Therefore the very 
past is related to the instantaneous value of $c(t)$. In the opposite case, at the lower limit 
$t^{\prime} \simeq 0$, the change of the concentration near to the initial value 
$\partial_{t^{\prime}}\,c(t^{\prime} = 0)$ is directly coupled to the instantaneous value $c(t)$. 
In such a manner the memory term represents a weighted coupling of the behaviour at the 
initial time and the observation time. This coupling leads to another long-time 
behaviour. Notice that the generic behaviour, discussed below, is not changed by assuming 
other terms with different power laws in Eq.~(\ref{ev4}). 

\subsection{Stationary solutions}

In this section we find the stationary solutions of Eq.~(\ref{ev4}) and discuss 
their stability. An important case is realized by $r > 0,~u > 0$ and arbitrary $\kappa$. 
Introducing dimensionless quantities $c \to r/u c,~t \to t/r$ we end up with an equation
\begin{equation}
\partial_t c(t) = c(t) - c^2(t) - 
\mu \int_0^t c(t - t^{\prime})~\partial_{t^{\prime}} c(t^{\prime}) dt^{\prime}
\label{ev5}
\end{equation}
with $\mu = \kappa/u$. The solution of the evolution Eq.~(\ref{ev5}) is simple when the memory kernel is 
zero, i.e. $\mu = 0$. It results
\begin{equation}
c(t) = \frac{c_0 e^{t}}{1 + c_0 (e^{ t}-1)}\quad \mbox{with}\quad c_0 = c(t=0)\quad.
\label{so}
\end{equation}
In the long time limit it results a non-trivial stationary and stable solution 
$c_s(\mu = 0) = 1$ and an unstable trivial solution $c_s = 0$. The inclusion of a memory 
term will change that behaviour drastically. To that aim let us analyse how the 
concentration is controlled by the feedback-coupling strength $\mu$. The formal 
solution of Eq.~(\ref{ev5}) is obtained by using Laplace transformation. We get
\begin{align} 
c(z) &=  \frac{c_0 (1 + \mu\, c(z))- A(z)}{z(1 + \mu\, c(z)) + 1}\nonumber\\
\mbox{with}\quad A(z) &= \mathcal{L}(c^2(t))(z)\quad.
\label{so1}
\end{align} 
Remark tat we discuss only the non-trivial memory controlled solution by assuming 
$c_0 \neq 0$ and $c_0 \neq 1$. Following the line of the previous section the 
long-time behaviour is obtained by making the ansatz $c(t) = c_s + \varphi (t)$ or 
after Laplace transformation 
\begin{equation}
c(z) = \frac{c_s}{z} +  \varphi (z)
\label{so2}
\end{equation}
where the function $\varphi (z)$ remains regular for $z \to 0$. This function will be discussed 
in the next subsection in combination with the linear stability. The quantity 
$c_s(\mu)$ represents, like before, the stationary solution in the limit $t \to \infty$. 
Apart from the trivial solution $c_s = 0$ there exists a non-trivial, memory controlled solution
\begin{equation}
c_s(\mu) = \frac{1 + \mu\, c_0}{1 + \mu} \quad.
\label{so3}
\end{equation}
Notice that this result is in accordancce with the general conclusion expressed by Eq.~(\ref{lt}). 
The stationary solution, Eq.~(\ref{so3}), depends on both, the initial value $c_0$ and the memory strength 
$\mu $. If one applies the Laplace transformation in Eq.~(\ref{disdel}) and making linear stability analysis, see 
the subsequent subsection one ends up only with the memory independent fixed points $c_s = 1$ (stable) and 
$c_s = 0$ (unstable), respectively. Here, we have demonstrated that an internal feedback leads to  a 
modified non-trivial stationary fixed point. Because $c(t)$ is a concentration the 
stationary solution is only accessible in case $c_s \geq 0$. The non-Markovian behaviour is apparent by 
the dependence of $c_s(\mu)$ on the initial concentration $c_0$ and on the memory parameter $\mu $.

\subsection{Phase diagram}

\begin{figure}[!ht]
\centering
\psfrag{p0}[][][1.5]{$c_0$}
\psfrag{mu}[][][1.5]{$\mu$}
\includegraphics{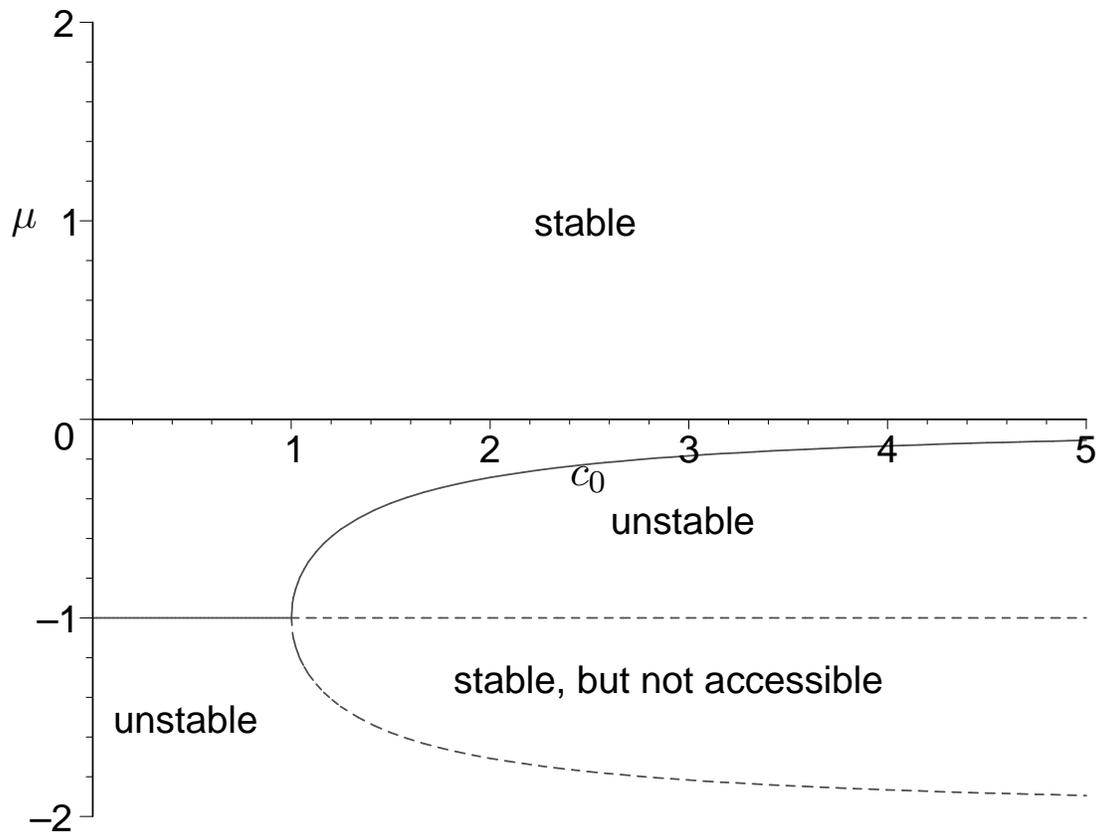}
\caption{Stability of the internal memory controlled stationary solution in the $c_0 - \mu$-plane.}
\label{Fig.3}
\end{figure}

In this subsection the phase diagram is obtained employing a linear stability 
analysis.  Notice that it can not be performed in terms of $\varphi (z) $ defined in Eq.~(\ref{so2}). 
Instead of that we have to insert $c(t) = c_s(\mu) + \varphi (t)$ in Eq.~(\ref{ev5}). As the result one finds  
$\varphi \propto \exp(-\Lambda t)$, where the stability exponent $\Lambda$ reads 
\begin{align}
\Lambda &= -1 \quad\quad\quad\quad\qquad~~~\mbox{if}\quad c_s = 0 \nonumber\\ 
\Lambda &= \frac{c_0\, \mu^2 + 2\, c_0 \mu +1}{1 + \mu}\quad\mbox{if}\quad c_s \neq 0 \quad.
\label{sta}
\end{align}
The phase diagram is defined in the $\mu - c_0 - $plane under exclusion of the trivial case $c_0 =1$. In the 
stable region both conditions, $c_s > 0$ and $\Lambda > 0$, has to be fulfilled simultaneously. Notice that 
there exits regions where the stationary solution is available $c_s \geq 0$ but not stable $\Lambda < 0$. The 
different cases are summarized in Fig.\ref{Fig.3}. The non-trivial stationary solution 
is stable for $\mu > -1$ in the interval $0 < c_0 \leq 1$ and for $c_0 \geq 1$ in case 
of $\mu \geq \mu_1\,(c_0) = - 1 + \sqrt{1 - c_0^{-1}}$. There exists a second domain  
where the non-trivial stationary solution should be stable. This area is limited by the lines 
$\mu = -1$ and $ \mu_2(c_0) = -1 - \sqrt{1 - c_0^{-1}} $ (dashed lines in Fig.\ref{Fig.3}). However the system can not 
achieve this region, because the initial value $c_0 > 1$ leads to $c_s(\mu ) > c_0$ in contradiction to the 
monotone decrease of the function $c(t)$ as indicated by Eq.~(\ref{ev5}). We omit the discussion of the other 
cases with different signs of the parameter $r$ and $u$ in Eq.~(\ref{ev4}). The inclusion of an internal memory 
leads to a non-trivial stationary point which is controlled strongly by the memory parameter $\mu $. Because the 
feedback is characterized by the concentration, relation Eq.~(\ref{lt}) is violated leading to a new fixed 
point in comparison to the non-Markovian situation.

\section{Conclusions}

In the present paper we have generalized the conventional rate equations of the reaction 
limited case by including memory effects. Once, the memory can be originated by external 
constraints which hints the particle to encounter simultaneously. As a consequence the rate of the 
concentration at the present time may be coupled to the rate at a previous time, where between 
both processes an additional delay-time $\tau$ appears. Although the stationary point are unchanged by 
such kind of external delay, the dynamical behaviour to reach that fixed points is altered. This situation 
reminds to an additive noise coupling in stochastic processes. In case of an internal feedback the stationary 
solution is changed and depends on the memory coupling and the initial value. The reason for that 
consists of the self-organized manner inherent in the memory. The time scale 
of the relevant variable, the concentration, determines the time scale of the feedback. Both realizations are 
characterized by an explicit coupling of the rates at the observation time $t$ and a previous time 
$0 \le t' \le t$ or with other words, the instantaneous changing rate of the concentration is not only determined 
by time-local gain and loss terms, but additionally by the changing rate in the past. 
Such a situation may realized whenever the reacting entities are not available 
simultaneously. Thus, the reaction is determined by the accumulation of the 
species at previous times. To capture the influence of that delay process, 
a memory term is included into the evolution equation. The non-Markovian part is 
generally found adopting projection methods of statistical mechanics. Especially, the 
form of the memory is suggested by analyzing glasses or anomalous diffusion in disordered media. 
Although we are aware, that the analytical expression of the feedback term is the most 
controversial point, the delay is in according to the general approach applied for 
chemical reactions. The memory term allows to study a real competitive situation 
between two non-linear terms in our evolution equation, namely the instantaneous loss term 
and the memory driven one. Even that kind of competition leads to the richer behaviour discussed 
in the present paper. Due to the memory the present and the initial times are mixed. Therefore, 
the system offers the possibility that the reaction process can be reduced or or speeded up. 
As the result we get a new stationary behaviour which differs significantly from the standard one. 
We believe that memory effects are also a feature of other dynamical complex 
systems.

\begin{acknowledgments} 
We thank Michael and Beatrix Schulz as well as Marian Brandau for fruitful discussions. 
This work was supported by DFG (SFB 418).
\end{acknowledgments}

\newpage

\end{document}